\documentclass[prd,superscriptaddress,amsfonts,amssymb,amsmath,showpacs,twocolumn]{revtex4-2}
\usepackage{bm}
\usepackage{amsfonts}
\usepackage{latexsym}
\usepackage[latin1]{inputenc}
\usepackage{graphicx}
\usepackage{amsmath}
\usepackage{palatino}
\usepackage{mathpazo}
\usepackage{textcomp}
\usepackage{float}
\usepackage{booktabs}
\usepackage{dcolumn}
\usepackage{ragged2e}
\usepackage{hyperref}
\hypersetup{colorlinks,citecolor=blue}
\hypersetup{colorlinks=true,linkcolor=red,filecolor=magenta,    urlcolor=blue}
\usepackage{amsmath}
\usepackage{xcolor}
\usepackage{orcidlink}
\usepackage[caption=false]{subfig}
\usepackage{commath}
\def\jnl@style{}
\def\aaref@jnl#1{{\jnl@style#1}}

\def\aaref@jnl#1{{\jnl@style#1}}

\def\aj{\aaref@jnl{AJ}}                   
\def\apj{\aaref@jnl{ApJ}}                 
\def\apjl{\aaref@jnl{ApJ}}                
\def\apjs{\aaref@jnl{ApJS}}               
\def\apss{\aaref@jnl{Ap\&SS}}             
\def\aap{\aaref@jnl{A\&A}}                
\def\aapr{\aaref@jnl{A\&A~Rev.}}          
\def\aaps{\aaref@jnl{A\&AS}}              
\def\mnras{\aaref@jnl{Mon.~Not.~Roy.~Astron.~Soc.}}             
\def\prd{\aaref@jnl{Phys.~Rev.~D}}        
\def\plb{\aaref@jnl{Phys.~Lett.~B}}        
\def\prc{\aaref@jnl{Phys.~Rev.~C}}  
\def\prl{\aaref@jnl{Phys.~Rev.~Lett.}}    
\def\qjras{\aaref@jnl{QJRAS}}             
\def\skytel{\aaref@jnl{S\&T}}             
\def\ssr{\aaref@jnl{Space~Sci.~Rev.}}     
\def\zap{\aaref@jnl{ZAp}}                 
\def\nat{\aaref@jnl{Nature}}              
\def\aplett{\aaref@jnl{Astrophys.~Lett.}} 
\def\apspr{\aaref@jnl{Astrophys.~Space~Phys.~Res.}} 
\def\physrep{\aaref@jnl{Phys.~Rep.}}      
\def\physscr{\aaref@jnl{Phys.~Scr}}       
\def\commat{\aaref@jnl{Comm.~Math.~Phys.}}              
\def\science{\aaref@jnl{Science}}               
\def\cqg{\aaref@jnl{Classical Quant.~Grav.}}            
\def\jpcs{\aaref@jnl{JPCS}}                                     
\def\ijmpd{\aaref@jnl{Int.~J.~Mod.~Phys.~D}}                    
\def\grg{\aaref@jnl{Gen.~Relat.~Gravit.}}               
\def\rpp{\aaref@jnl{Rep.~Prog.~Phys.}}          
\def\npa{\aaref@jnl{Nucl.~Phys.~A}}        
\def\lrr{\aaref@jnl{Living Rev.~Rel.}}                   
\def\jcap{\aaref@jnl{J.~Cosmology Astropart.~Phys.}}    
\def\rmp{\aaref@jnl{Rev.~Mod.~Phys.}}   
\def\epjc{\aaref@jnl{Eur.~Phys.~J.~C}}

\allowdisplaybreaks[1]
\renewcommand{\arraystretch}{1.1}
\begin{document}
\color{black}       

\title{Constraint on the  equation of state parameter ($\omega$) in non-minimally coupled $f(Q)$ gravity}

\author{Sanjay Mandal\orcidlink{0000-0003-2570-2335}}
\email{sanjaymandal960@gmail.com}
\affiliation{Department of Mathematics, Birla Institute of Technology and
Science-Pilani,\\ Hyderabad Campus, Hyderabad-500078, India.}


\author{P.K. Sahoo\orcidlink{0000-0003-2130-8832}}
\email{pksahoo@hyderabad.bits-pilani.ac.in}
\affiliation{Department of Mathematics, Birla Institute of Technology and
Science-Pilani,\\ Hyderabad Campus, Hyderabad-500078, India.}
%
\date{\today}
\begin{abstract}

We study observational constraints on the modified symmetric teleparallel gravity, the non-metricity $f(Q)$ gravity, which reproduces background expansion of the universe. For this purpose, we use Hubble measurements, Baryonic Acoustic Oscillations (BAO), 1048 Pantheon supernovae type Ia data sample which integrate SuperNova Legacy Survey (SNLS), Sloan Digital Sky Survey (SDSS), Hubble Space Telescope (HST) survey, Panoramic Survey Telescope and Rapid Response System (Pan-STARRS1). We confront our cosmological model against observational samples to set constraints on the parameters using Markov Chain Monte Carlo (MCMC) methods. We find the equation of state parameter $\omega=-0.853^{+0.015}_{-0.020}$ and $\omega= -0.796^{+0.049}_{-0.074}$ for Hubble and Pantheon samples, respectively. As a result, the $f(Q)$ model shows the quintessence behavior and deviates from $\Lambda$CDM.

\end{abstract}


\maketitle

\section{Introduction}

The current cosmic expansion of the universe motivates the scientific community to understand its fundamental properties. The present status of the universe can comprehend by testing the $\Lambda$ cold dark matter ($\Lambda$CDM) model, cosmological models, and any deviation from it. Moreover, General Relativity (GR) is a well-established gravitational theory, but its extensions and modifications are seeking more interest due to their successful properties in describing the universe's accelerated expansion \cite{Capo/2011, Sari}. Hence, we aim to develop gravitational theories which have GR as a specific limit. But, in general, that formulation must include extra degree(s) of freedom to fulfill the above prerequisites.

This report will highlight the cosmological model based on the recently proposed extension of symmetric teleparallel gravity, so-called $f(Q)$ gravity, where the non-metricity $Q$ describes the gravitational interaction \cite{Jimenez/2018}. Investigations on $f(Q)$ gravity have developed rapidly and lead to interesting applications \cite{Lazkoz/2019, Mandal/2020, Mandal/2020a, Harko/2018, Barros/2020, Jimenez/2020, Hasan/2021, Solanki/2021, Frus/2021, Flat/2021, Khyllep/2021,Anag/2021, Ata/2021,Dial/2019,Ayu/2021, Amb/2020}. If we look at the universe's expansion history, one can see that some cosmological parameters play an important role in designating the cosmological model's cosmic evolution. And, it is well known that the equation of state parameter $(\omega)$ predicts various fluid descriptions of the universe. Therefore, it is interesting to constraint this parameter using observational data. For this purpose, we use Hubble measurements, Baryonic Acoustic Oscillations (BAO), Pantheon supernovae type Ia integrates SuperNova Legacy Survey (SNLS), Sloan Digital Sky Survey (SDSS), Hubble Space Telescope (HST) survey, Panoramic Survey Telescope and Rapid Response System (Pan-STARRS1). The MCMC methods use to do the numerically analysis.

In this report, the ideas presented in the following sections. In section \ref{sec2}, we discuss the basic setup for the $f(Q)$ gravity. In section \ref{sec3}, we discuss the cosmological application in FRW space-time. In section \ref{sec4}, we discuss the various type of observational datasets, constraint the parameters using the MCMC method, and deliberate our results. Finally, gathering all the information, we conclude in section \ref{sec5}.

\section{Action and Field equations}\label{sec2}

Here, we consider the action for matter coupling in $f(Q)$ gravity is given by \cite{Jimenez/2018}
\begin{equation}
\label{1}
S=\int d^4x \sqrt{-g}\left[\frac{1}{2}f_1(Q)+f_2(Q)L_M\right],
\end{equation}
where $g$ is the determinant of metric, $f_1(Q) \,\& \, f_2(Q)$ are the arbitrary functions of the non-metricity $Q$, and $L_M$ is the Lagrangian for the matter fields.

The nonmetricity tensor and its traces are such that
\begin{equation}
\label{2}
Q_{\gamma\mu\nu}=\nabla_{\gamma}g_{\mu\nu}\,,
\end{equation}
\begin{equation}
\label{3}
Q_{\gamma}={{Q_{\gamma}}^{\mu}}_{\mu}\,, \qquad \widetilde{Q}_{\gamma}={Q^{\mu}}_{\gamma\mu}\,.
\end{equation}
Moreover, the superpotential as a function of nonmetricity tensor is given by
\begin{equation}
\label{4}
4{P^{\gamma}}_{\mu\nu}=-{Q^{\gamma}}_{\mu\nu}+2Q_{({\mu^{^{\gamma}}}{\nu})}-Q^{\gamma}g_{\mu\nu}-\widetilde{Q}^{\gamma}g_{\mu\nu}-\delta^{\gamma}_{{(\gamma^{^{Q}}}\nu)}\,,
\end{equation}
where the trace of nonmetricity tensor \cite{Jimenez/2018} reads
\begin{equation}
\label{5}
Q=-Q_{\gamma\mu\nu}P^{\gamma\mu\nu}\,.
\end{equation}
To simplify the formulation, let us introduce the following notations
\begin{align}
\label{6}
f=f_1(Q)\,+2 f_2(Q)L_M,\\
F=f_1'(Q)+2 f_2'(Q)L_M,
\end{align}
where primes (') represent the derivatives of functions $f_1(Q) \,\& \, f_2(Q)$ with respect to $Q$.

The energy-momentum tensor for the fluid description of the spacetime can be written by its definition
\begin{equation}
\label{6}
T_{\mu\nu}=-\frac{2}{\sqrt{-g}}\frac{\delta(\sqrt{-g}\mathcal{L}_m)}{\delta g^{\mu\nu}}\,.
\end{equation}
By varying action \eqref{1} with respect to metric tensor, we can write the gravitational field equation, which is given by
\begin{multline}
\label{7}
\frac{2}{\sqrt{-g}}\nabla_{\gamma}\left( \sqrt{-g}F {P^{\gamma}}_{\mu\nu}\right)+\frac{1}{2}g_{\mu\nu}f_1 \\
+F \left(P_{\mu\gamma i}{Q_{\nu}}^{\gamma i}-2Q_{\gamma i \mu}{P^{\gamma i}}_{\nu} \right)=-f_2 T_{\mu\nu}\,.
\end{multline}
Now, one can use \eqref{7} to explore the cosmological applications in $f(Q)$ modified gravity.

\section{The $f(Q)$ cosmology}\label{sec3}

To explore several cosmological applications, we presume the homogeneous, isotropic and spatially flat line element given by
\begin{equation}
\label{8}
ds^2 = -N^2(t) dt^2 +a^2(t) \delta_{ij} dx^i dx^j,
\end{equation}
where $N(t)$ is the lapse function and for the usual time reparametrization freedom, we can take $N=1$ at any time. $\delta_{ij}$ is the Kronecker delta and $i,\,\, j$ run over spatial components. The expansion rate and dilation rate can be written as
\begin{equation}
\label{9}
H=\frac{\dot{a}}{a},\,\,\,\,\ T=\frac{\dot{N}}{N},
\end{equation}
respectively. For this line element the non-metricity read as $Q=6 (H/N)^2$.

We shall work on the perfect fluid matter distribution, for which the energy momentum tensor \eqref{6} become diagonal. The gravitational equations \eqref{7} in this case generalized to two Friedman equations:
\begin{equation}
\label{10}
f_2 \rho =\frac{f_1}{2}-6 F \frac{H^2}{N^2},
\end{equation}
\begin{equation}
\label{11}
-f_2 p = \frac{f_1}{2}-\frac{2}{N^2}[(\dot{F}-FT)H+F(\dot{H}+3H^2)],
\end{equation}
respectively. Here, $\rho$ is the energy density and $p$ is the pressure of the fluid content of the spacetime. It is easy to verify that, for $f_1=-Q$ and $f_2=1=-F$, the above Friedman equations reduce to standard one \cite{Harko/2018}. From the above field equations, the continuity equation for matter field can be derived as
\begin{equation}
\label{12}
\dot{\rho} +3 H(\rho+p) = -\frac{6 f_2'H}{f_2N^2}(\dot{H}-HT)(L_M+\rho).
\end{equation}
From \eqref{12}, one can recover the standard continuity equation by imposing $L_M=-\rho$ as
\begin{equation}
\label{13}
\dot{\rho} +3 H(\rho+p) =0
\end{equation}
This is compatible with the isotropic and homogeneous background of the universe (see more details about the continuity equation in $f(Q)$ gravity \cite{Harko/2018}).

\subsection{Cosmological Model}\label{sub1}

In this subsection, we shall proceed with $N=1$. Since, we are working on the Friedman-Robertson-Walker (FRM) framework, the non-metricity $Q$ and dilation rate $T$ reduce to
\begin{equation}
\label{14}
Q=6H^2,\,\,\,\, T=0.
\end{equation}
Moreover, the modified Friedman equations \eqref{10} and \eqref{11} can be rewritten as
\begin{equation}
\label{15}
3H^2= \frac{f_2}{2 F}\left(-\rho+\frac{f_1}{2f_2}\right),
\end{equation}
\begin{equation}
\label{16}
\dot{H}+3H^2+\frac{\dot{F}}{F}H = \frac{f_2}{2 F}\left(p+\frac{f_1}{2f_2}\right).
\end{equation}
Now, we have two equations with five unknown such as $H,\,\, \rho,\,\, p,\,\, f_1,\,\, f_2$. To proceed further, we consider the relation $p=\omega \rho$ and from equation \eqref{13}, we find the following relation
\begin{equation}
\label{17}
\rho =\rho_0 a(t)^{-3 (1+\omega)},
\end{equation}
where $\rho_0$ is the proportionality constant and $\omega$ is the equation of state parameter. The relation between scale factor $a(t)$ and redshift $z$ read 
\begin{equation}
\label{18}
a(t)=\frac{a_0}{1+z}.
\end{equation}
Without loss of generality, we adopt $\rho_0=a_0=1$. In modified theories of gravity the cosmological scenarios can be discussed through the properties of cosmological models. For this purpose, we introduce two Lagrangian functions $f_1(Q)$ and $f_2(Q)$ as
\begin{align}
\label{19}
f_1(Q)=\alpha Q^n,\,\,\,\   f_2(Q)= Q,
\end{align}
where $\alpha$ and $n\neq 1$ are the arbitrary constants.

Using \eqref{17}, \eqref{18}, \eqref{19} in \eqref{15}, we find the following expression for $H(z)$
\begin{equation}
\label{20}
H(z)=\left\lbrace\frac{2\,(1+z)^{3(1+\omega)}}{\alpha\, (2n-1) \,6^{n-1}}\right\rbrace^{\frac{1}{2n-2}}.
\end{equation}
Now, our aim is to put constraint on the parameters $\alpha,\,\, n,\,\, \omega$ using the astronomical observation data.

\section{Data, Methodology and Results}\label{sec4}

This section deals with the various observational datasets to constraint the parameters $\alpha,\,\, n,\,\, \omega$. For this, we have adopted some statistical analysis to perform the numerical analysis. Specifically, we employ a Markov Chain Monte Carlo (MCMC) method to obtain the posterior distributions of the parameters, using the standard Bayesian technique. This stimulation is done by using the Hubble measurements ( i.e., Hubble data) and SNe Ia data. The best fits of the parameetrs are maximized by using the probability function
\begin{equation}
\label{21}
\mathcal{L} \propto exp(-\chi ^2/2),
\end{equation}
where $\chi^2$ is the \textit{pseudo chi-squared function} \cite{baye}. The $\chi^2$ functions for various dataset are discussed below.

\subsection{Hubble Dataset}
Recently, a list of 57 data points of Hubble parameter in the redshift range $0.07\leq z\leq 2.41$ were compiled by Sharov and Vasiliev \cite{Sharov/2018}. This H(z) dataset was measured from the line-of-sight BAO data \cite{BAO1, BAO2, BAO3, BAO4, BAO5} and the differential ages $\Delta t$ of galaxies \cite{h1,h2,h3,h4}. The complete list of datasets is presented in \cite{Sharov/2018}. To estimate the model parameters, we use the Chi-square function which is given by
\begin{equation}
\label{22}
\chi^2_{OHD}(p_s)=\sum_{i=1}^{57}\frac{[H_{th}(p_s,z_i)-H_{obs}]^2}{\sigma^2_{H(z_i)}},
\end{equation}
where $H_{obs}(z_i)$ represents the observed Hubble parameter values, $H_{th}(p_s,z_i)$ represents the Hubble parameter with the model parameters, and $\sigma^2_{H(z_i)}$ is the standard deviation.
In Fig. \ref{f1}, the profile of our model against Hubble data shown. The marginalized constraining results are displayed in Fig. \ref{f3}.

\begin{widetext}

\begin{figure}[H]
\includegraphics[scale=0.9]{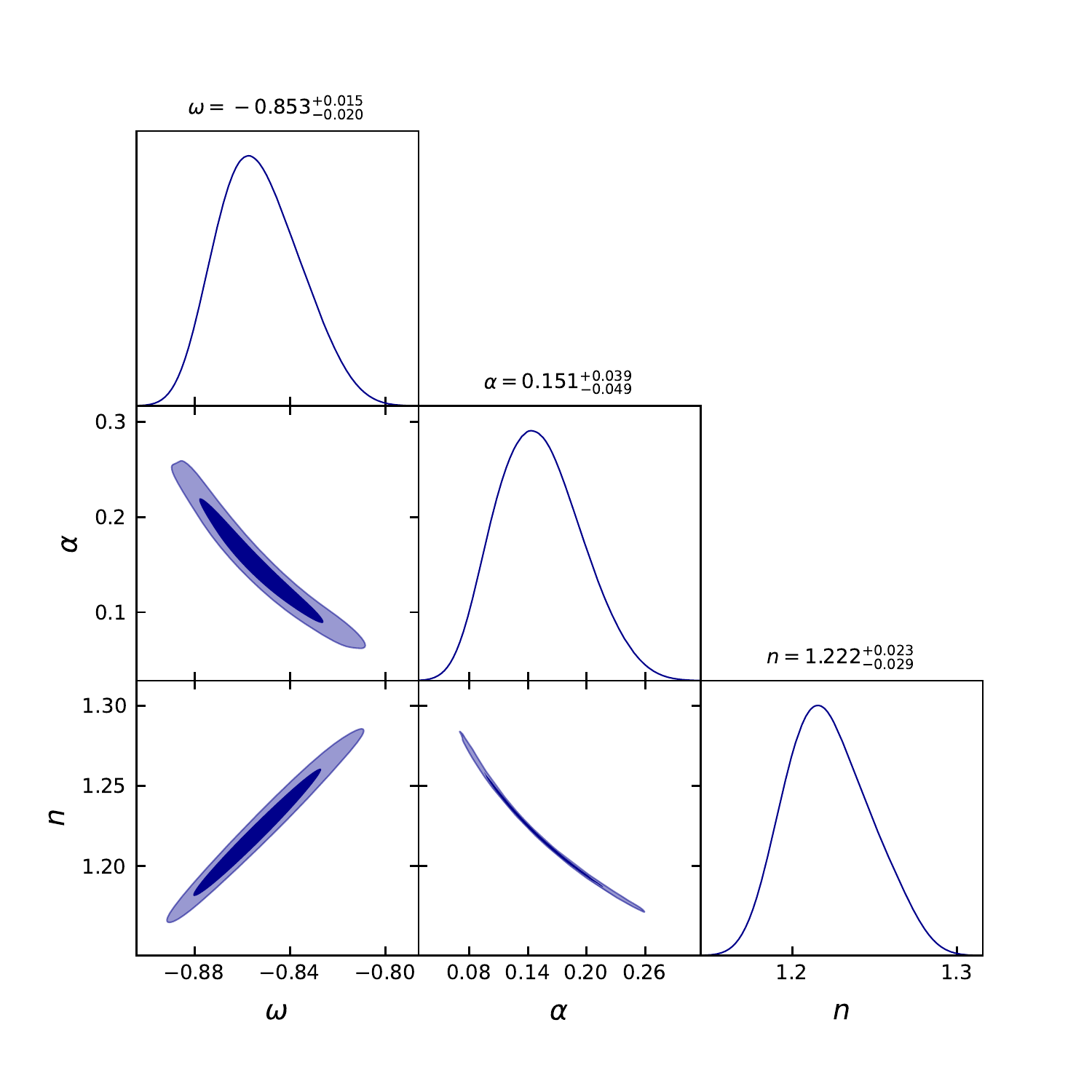}
\caption{The marginalized constraints on the parameters included in the expression of Hubble parameter $H(z)$ (i.e., in Eqn. \eqref{20}) are presented by using the Hubble sample.}
\label{f2}
\end{figure}
\end{widetext}

\subsection{Pantheon Dataset}

Supernovae type Ia is a powerful distance indicator to explore the background evolution of the universe. Therefore, to constraint the above parameters, we take the recent Pantheon supernovae type Ia sample, which is collectively 1048 SNe Ia data points from various SN Ia samples in the redshift-range $z \in [0.01,2.3]$ such as SDSS, SNLS, Pan-STARRS1, low-redshift survey, and HST surveys \cite{Scolnic/2018}. The $\chi^2_{SN}$ function is given to do the statistical analysis for this sample \cite{Scolnic/2018} by 
\begin{equation}
\chi^2_{SN}(p_1,....)=\sum_{i,j=1}^{1048}\bigtriangledown\mu_{i}\left(C^{-1}_{SN}\right)_{ij}\bigtriangledown\mu_{j},
\end{equation}
where $p_j$ represents the assumed model's free parameters and $C_{SN}$ is the covariance metric \cite{Scolnic/2018}, and $\mu$ represents the distance moduli is given by;
 \begin{align*}
 \mu^{th}(z)& =5\log\frac{D_L(z)}{10pc},\, \quad D_L(z)=(1+z)D_M,\\
 D_M(z)&=c \int_0^{z}\frac{d\tilde{z}}{H(\tilde{z})},\, \quad \bigtriangledown\mu_{i}=\mu^{th}(z_i,p_1,...)-\mu_i^{obs}.
 \end{align*}
 
\begin{widetext}

\begin{figure}
\includegraphics[scale=0.65]{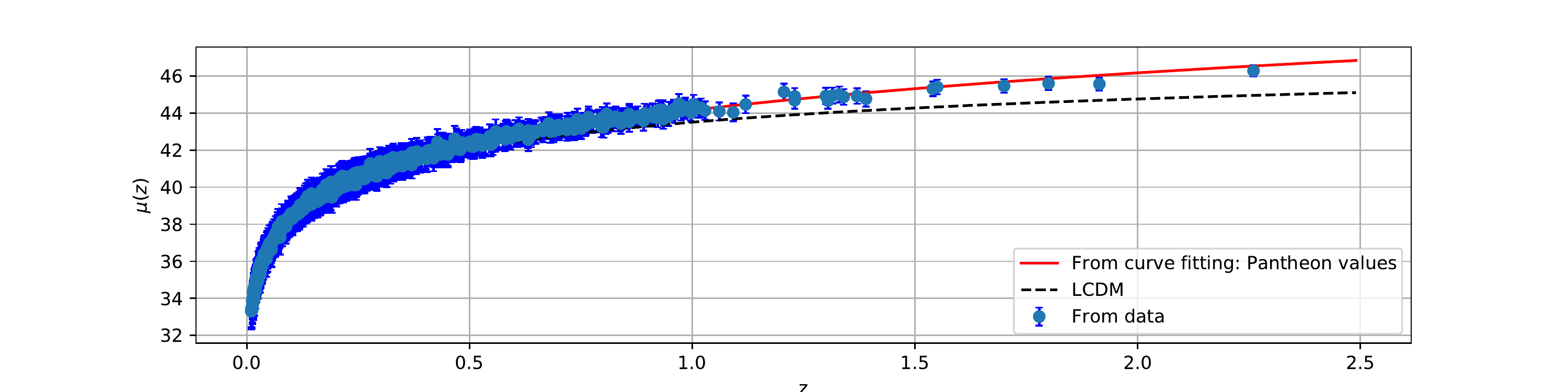}
\caption{The evolution of $\mu(z)$ with respect to redshift $z$ is shown here. The red line represents our model and dashed-line indicates the $\Lambda$CMD model with $\Omega_{m0}=0.3$ and $\Omega_{\Lambda 0}=0.7$. The dots are shown the 1048 Pantheon dataset with error bar. }
\label{f3}
\end{figure}

\end{widetext}

 \begin{widetext}
 
\begin{figure}[H]

\includegraphics[scale=0.5]{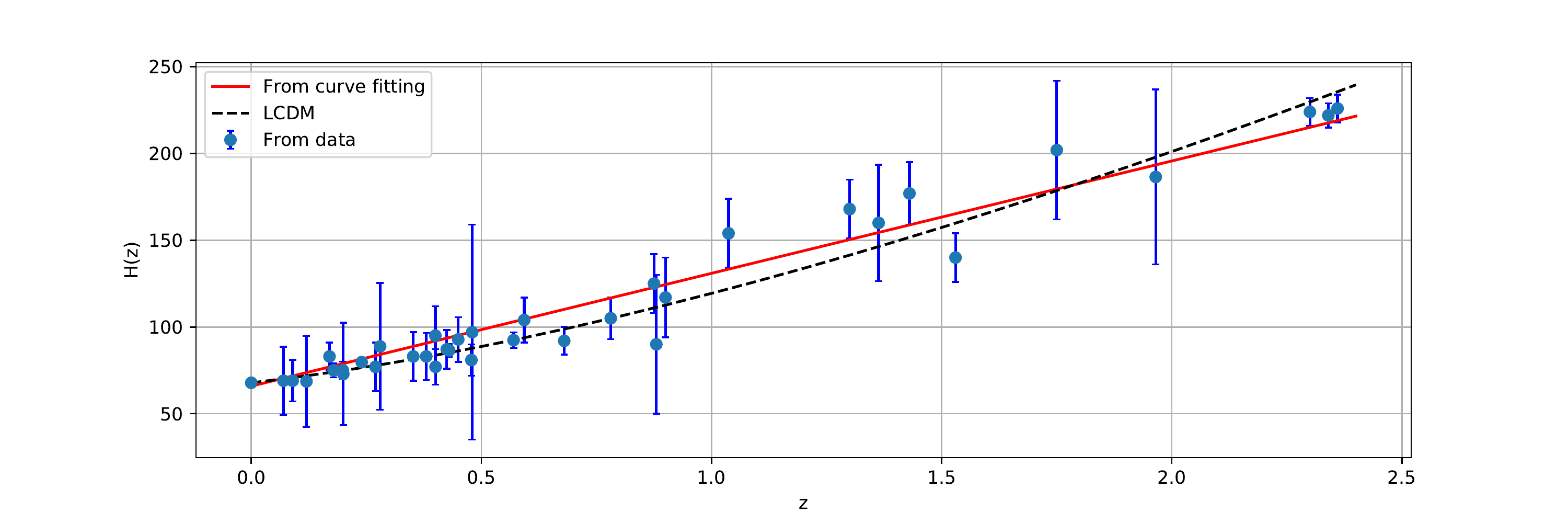}
\caption{The evolution of Hubble parameter $H(z)$ with respect to redshift $z$ is shown here. The red line represents our model and dashed-line indicates the $\Lambda$CMD model with $\Omega_{m0}=0.3$ and $\Omega_{\Lambda 0}=0.7$. The dots are shown the Hubble dataset with error bar.}
\label{f1}

\end{figure}

Figure \ref{f3} represent the best fit of our model against the Pantheon dataset and the posterior distributions of the parameters are shown in Figure \ref{f4}.

\begin{figure}[H]
\includegraphics[scale=0.9]{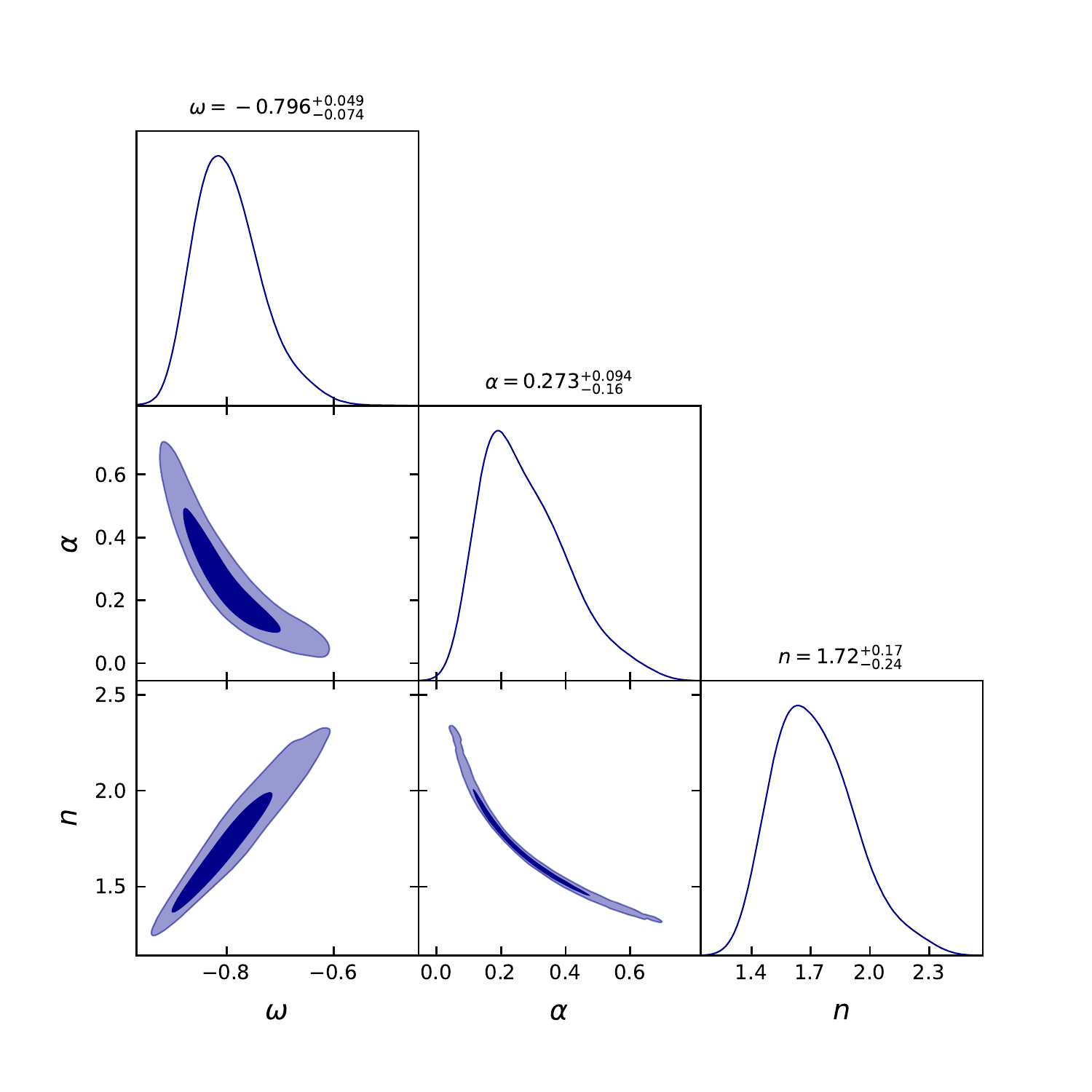}
\caption{The marginalized constraints on the parameters included in the expression of Hubble parameter $H(z)$ (i.e., in Eqn. \eqref{20}) are presented by using the Pantheon sample.}
\label{f4}
\end{figure}
\end{widetext}

\subsection{Results}

In Table \ref{t1}, we show constraints at $68\%$ C.L. of the cosmological parameter $\omega$ and model parameters $\alpha, n$ for the cosmological model. Figures \ref{f2}, \ref{f4} present the contour plots of the parameters at $68\%$ and $95\%$ C.L.  for Hubble and Pantheon samples, respectively. For reference, we compare the $f(Q)$ model with the $\Lambda$CDM model with constraints values of parameters in Figures \ref{f1} and \ref{f2}. It is observed that the $f(Q)$ model is perfectly fitting with the observational data and deviates slightly from the $\Lambda$CDM. Moreover, it is well-known that the present scenario of the universe, i.e., accelerated expansion, can be discussed with the presence of additional constant $\Lambda$ in Einstein's field equations or by modifying the fundamental formulation of gravity for the evolution of the universe. Besides this, the equation of state parameter $(\omega)$ plays a vital role for a cosmological model to predict its' different phases of evolution. The present scenario of the universe can predict by either quintessence behavior of $\omega\,\, (i.e., -1<\omega<-1/3)$ or phantom behavior of $\omega\,\, (i.e., \omega<-1)$. For our model, we found $\omega=-0.853^{+0.015}_{-0.020}$ for Hubble dataset and $\omega=-0.796^{+0.049}_{-0.074}$ for Pantheon dataset at $1\sigma$ confidence level. One can clearly see that the $f(Q)$ model shows the quintessence behavior and it is near to $\Lambda$CDM.

\begin{table*}[!t]
	\renewcommand\arraystretch{1.5}
	\caption{The marginalized constraining results on three parameters $\omega,\,\, \alpha,\,\, n$ are shown by using the Hubble and Pantheon SNe Ia sample. We quote $1\,\sigma$ (68$\%$) errors for all the parameters here.
	}
	\begin{tabular} { l |c| c |c }
		\hline
		\hline

		Dataset    & $\omega$      &$\alpha$      &$n$        \\
		\hline
		Hubble   & $-0.853^{+0.015}_{-0.020}$   & $0.151^{+0.039}_{-0.049}$ &$1.222^{+0.023}_{-0.029}$  \\
		\hline
		Pantheon & $-0.796^{+0.049}_{-0.074}$   & $0.273^{0.094}_{-0.16}$ & $1.72^{+0.17}_{-0.24}$  \\
		
	    \hline
		\hline
	\end{tabular}
	\label{t1}
\end{table*}

\section{Conclusion}\label{sec5}

The rising concern in the current scenarios of the universe motivates us to go beyond the standard formulation of gravity. In this context, we have worked on the modified $f(Q)$ gravity to obtain observational constraints for the background candidates of the accelerated expansion of the universe. For this, we have used a wide variety of observational samples such as Hubble data and Pantheon data (which includes SDSS, SNLS, Pan-STARRS1, low-redshift survey, and HST surveys). The well-known parametrization technique is adopted to obtain the expression for $H(z)$. The best fit ranges of the parameters are obtained by applying the Bayesian method in MCMC simulation. The constraint values of the equation of state parameter $(\omega)$ suggest that the $f(Q)$ model shows quintessence behavior. In addition, we have depicted the profiles of Hubble parameter with the constraint values of parameters for both the dataset in Figures \ref{f1} and \ref{f3}, which helps us to compare our model with the $\Lambda$CDM. 

Moreover, one can compare our outcomes with the existing results to discuss the current scenario of the universe. Particularly, the cosmographic tool and observational constraints are widely used to discuss such scenarios in modified theories of gravity.  For instance, one can see some of the interesting studies used these ideas, such as the cosmographic idea is used to test the observational viability of various modified theories to discuss the universe's evolution. For example, the deceleration parameter $q$ is used to check the universe's status \cite{capo/2014}, and with the present values of cosmographic parameter are used checked the viability of class of $f(T)$ gravity \cite{capo/2011}, class of $f(Q)$ gravity \cite{Mandal/2020a}. Without a prior assumption on the equation of state parameter $\omega$ and with the cosmographic parameters the cosmological evolution studied through the profile of $\omega$ in $f(R)$ and $f(T)$ gravity \cite{capo/2019}. Also, some of the observational studies in $f(Q)$ gravity have been done in last few years \cite{Lazkoz/2019,Ata/2021,Anag/2021,Ayu/2021}. In most of these studies, authors have presumed different values of the equation of state parameter $\omega$ such as $\omega=1/3,\, \omega=0$ to include the radiation and matter content in the fluid description of the spacetime, respectively. Whereas in our study, such assumptions on $\omega$ are avoided, and the value of $\omega$ constraint against the observational data. And, this is the advantage of our work over such types of investigations. The main goal to describe the accelerated expansion of the universe is successfully archived.

In conclusion, our findings could motivate further research into the $f(Q)$ gravity because it is one of the alternatives to the coherence model that, aside from being preferred by the data. The significance of this model is that it does not face the cosmological constant problem because it does not comprise any additional constant inside the Lagrangian $f(Q)$ form. In a further study, it would be interesting to explore these types of models using weak lensing data, full CMB and LSS spectra, and other datasets. Some of these tests will be addressed in the near future, and we hope to report on them.

\section*{Acknowledgements}

S. Mandal acknowledges Department of Science and Technology (DST), Government of India, New Delhi, for awarding INSPIRE Fellowship (File No. DST/INSPIRE Fellowship/2018/IF180676). P.K. Sahoo acknowledges CSIR, New  Delhi, India for financial support to carry out the Research project [No.03(1454)/19/EMR-II, Dt. 02/08/2019]. We are very much grateful to the honorable referee and to the editor for the illuminating suggestions that have significantly improved our work in terms of research quality, and presentation.

\end{document}